\documentclass[twocolumn,letter]{jpsj2} %% for letters
\title{
Observation of Magnetic Monopoles in Spin Ice
}

\author{
Hiroaki \textsc{Kadowaki}$^{1}$, %\thanks{E-mail: kadowaki@phys.metro-u.ac.jp}, 
Naohiro \textsc{Doi}$^{1}$, 
Yuji \textsc{Aoki}$^{1}$, 
Yoshikazu \textsc{Tabata}$^{2}$, 
Taku J. \textsc{Sato}$^{3}$, 
Jeffrey W. \textsc{Lynn}$^{4}$, 
Kazuyuki \textsc{Matsuhira}$^{5}$, 
and 
Zenji \textsc{Hiroi}$^{6}$}

\inst{
$^{1}$Department of Physics, Tokyo Metropolitan University, 
Hachioji-shi, Tokyo 192-0397 \\
$^{2}$Department of Materials Science and Engineering, Kyoto University, 
Kyoto 606-8501 \\
$^{3}$NSL, Institute for Solid State Physics, University of Tokyo, 
Tokai, Ibaraki 319-1106 \\
$^{4}$NCNR, National Institute of Standards and Technology, 
Gaithersburg, Maryland 20899-6102, USA \\
$^{5}$Department of Electronics, Faculty of Engineering, Kyushu Institute of Technology, 
Kitakyushu 804-8550 \\
$^{6}$Institute for Solid State Physics, University of Tokyo, 
Kashiwa 277-8581}

\recdate{August 11, 2009; accepted September 8, 2009; published October 13, 2009}

\abst{
Excitations from a strongly frustrated system, 
the kagom\'{e} ice state of the spin ice Dy$_2$Ti$_2$O$_7$ 
under magnetic fields along a [111] direction, have been studied. 
They are theoretically proposed to be regarded as magnetic monopoles. 
Neutron scattering measurements of spin correlations show that close to 
the critical point the monopoles are fluctuating between high- and 
low-density states, supporting that the magnetic Coulomb force acts 
between them. 
Specific heat measurements show that monopole-pair creation obeys an 
Arrhenius law, indicating that the density of monopoles can be controlled 
by temperature and magnetic field.
}

\kword{magnetic monopole, spin ice, kagom\'{e} ice, neutron scattering, specific heat}

\begin{document}
\maketitle

Since the quantum mechanical hypothesis of the existence of 
magnetic monopoles proposed by Dirac \cite{Dirac31,Jackson}, 
many experimental searches have been performed, 
ranging from a monopole search in rocks of the moon to experiments 
using high energy accelerators \cite{Milton06}. 
But none of them was successful, and the monopole is an open question 
in high energy physics. 
Recently, theoretical attention has turned to 
condensed matter systems where tractable analogs of 
magnetic monopoles might be found \cite{Fang03,Castelnovo08,Qi09}, 
and one prediction\cite{Castelnovo08} is for an emergent elementary 
excitation in the spin ice \cite{Harris_HTO,Bramwell01} compound Dy$_2$Ti$_2$O$_7$. 

\begin{figure}[tb]
\begin{center}
\includegraphics[width=4.0cm,clip]{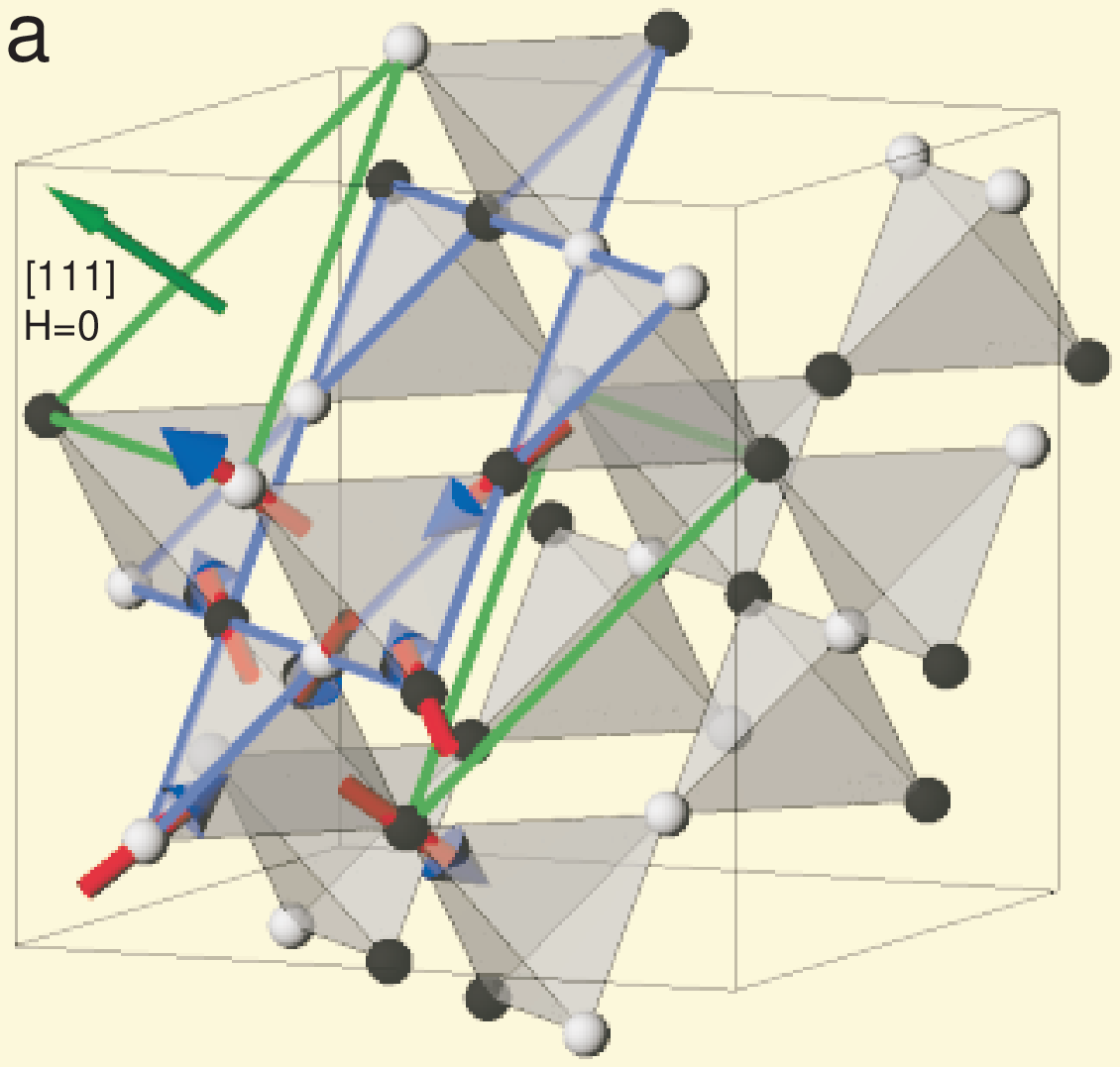}
\includegraphics[width=7.0cm,clip]{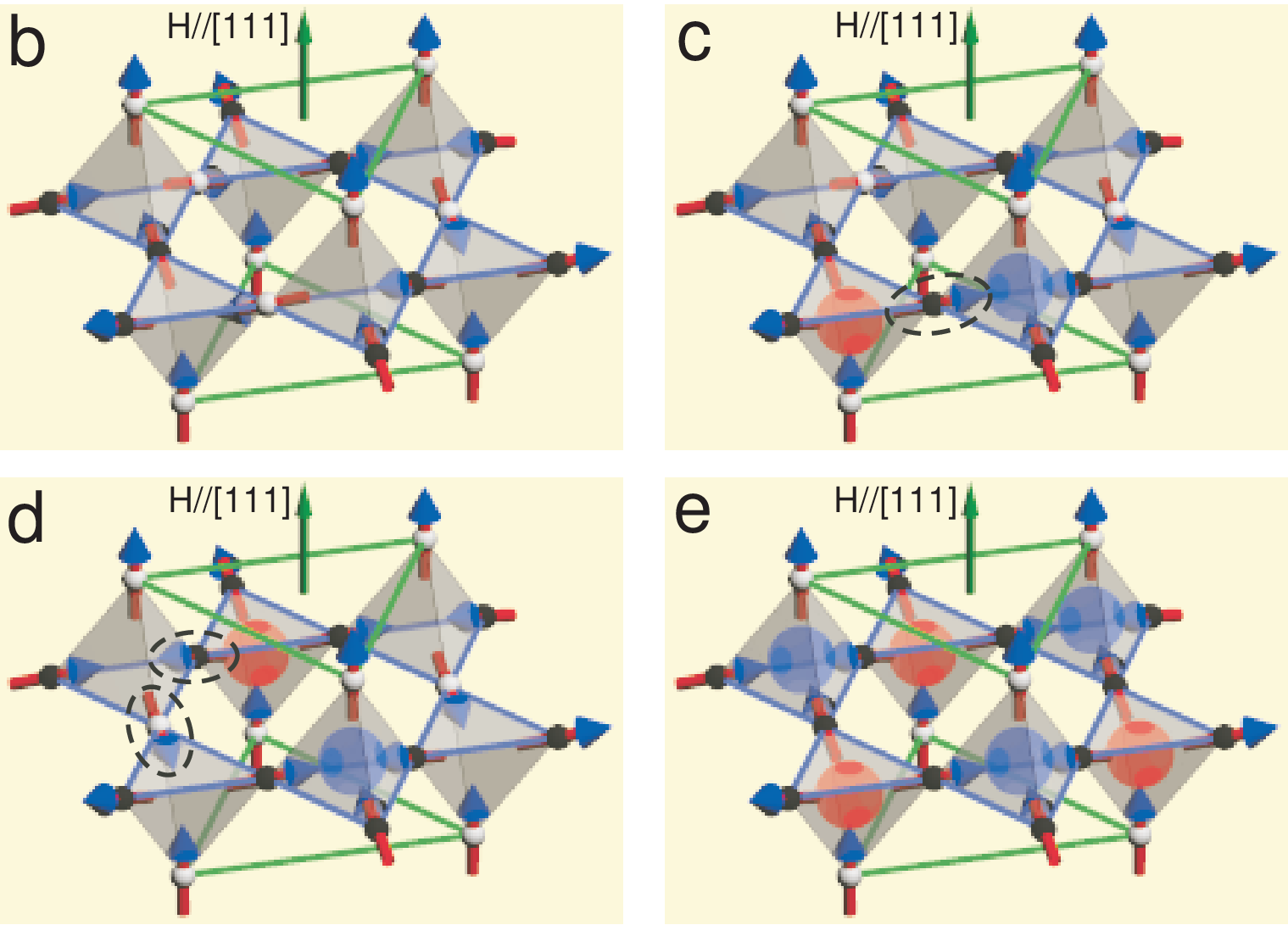}
\end{center}
\caption{
(Color) 
Magnetic moments of Dy$_2$Ti$_2$O$_7$ reside on the pyrochlore 
lattice \cite{Bramwell01}. 
At low temperatures, four magnetic moments on each tetrahedron obey 
the ice rule (2-in, 2-out). 
The resulting spin ice state is shown in (a). 
The pyrochlore lattice consists of stacked triangular and kagom\'{e} lattices, 
shown by green and blue lines, respectively, along a [111] direction. 
(b) Under small [111] magnetic fields, spins on the kagom\'{e} lattice 
remain in the disordered kagom\'{e} ice state \cite{Matsuhira_kagome}. 
(c) An excited state is induced by flipping a spin from (b), 
enclosed by a dashed circle, where neighboring tetrahedra have 
3-in, 1-out and 1-in, 3-out configurations. 
These ice-rule-breaking tetrahedra are represented by magnetic monopoles 
with opposite charges depicted by spheres. 
(d) By consecutively flipping two spins from (c), the monopoles 
are fractionalized. 
(e) As the magnetic field is increased, $H \gg H_{\mathrm{c}}$, 
spins realize a fully ordered, staggered arrangement of monopoles.
}
\label{f1}
\end{figure}

In solid water, the protons are disordered even at absolute zero 
temperature and thus retain finite entropy \cite{Pauling1935}, 
and spin ice exhibits the same type of disordered ground 
states \cite{Bramwell01,Ramirez_DTO}. 
The Dy spins occupy a cubic pyrochlore lattice, which is a 
corner sharing network of tetrahedra (Fig.~\ref{f1}(a)). 
Each spin is parallel to a local [111] easy axis, and interacts 
with neighboring spins via an effective ferromagnetic coupling.
This brings about a geometrical frustration where the lowest energy 
spin configurations on each tetrahedron follow the ice rule, 
``2-in, 2-out'' structure, and the ground states of the 
entire tetrahedral network are macroscopically degenerate in 
the same way as the disordered protons in water ice \cite{Pauling1935,Ramirez_DTO}.
In addition to this remarkable observation, there is the 
more intriguing possibility \cite{Castelnovo08} that the excitations from 
these highly degenerate ground states are topological in 
nature and mathematically equivalent to magnetic monopoles.

The macroscopic degeneracy of the spin ice state can be 
partly lifted by 
applying a small magnetic field along a [111] direction \cite{Matsuhira_kagome}. 
Along this direction the pyrochlore lattice consists of a stacking of triangular 
and kagom\'{e} lattices (Fig.~\ref{f1}(a)). 
In this field-induced ground state, the spins on the triangular lattices are 
parallel to the field and consequently drop out of the problem, while those on 
the kagom\'{e} lattices retain disorder under the same ice 
rules, only with a smaller zero-point entropy \cite{Moessner03}. 
This is referred to as the kagom\'{e} ice state 
\cite{sakakibara03,Matsuhira_kagome,Tabata_kagome,Higashinaka_kagome} 
(Fig.~\ref{f1}(b)).

In Fig.~\ref{f1}(c) we illustrate creation of a magnetic monopole and 
antimonopole pair in the kagom\'{e} ice state. 
An excitation is generated by flipping a spin 
on the kagom\'{e} lattice, which results in ice-rule-breaking 
``3-in, 1-out'' (magnetic monopole) and ``1-in, 3-out'' 
(anti-monopole) tetrahedral neighbors. 
From the viewpoint of the dumbbell model \cite{Castelnovo08}, where a magnetic 
moment is replaced by a pair of magnetic charges, the 
ice-rule-breaking tetrahedra simulate magnetic monopoles, 
with net positive and negative charges sitting on the centers of tetrahedra. 
The monopoles should interact via the magnetic Coulomb force \cite{Castelnovo08}, 
which is brought about by the dipolar interaction \cite{Hertog_mc} 
between spins in Dy$_2$Ti$_2$O$_7$. 
They can move and separate by consecutively flipping spins, 
but are confined to the two-dimensional kagom\'{e} layer 
(e.g. Fig.~\ref{f1}(d)). 
This possibility of separating the local excitation into its 
constituent parts is a novel fractionalization in a frustrated 
system in two or three dimensions \cite{Castelnovo08,Fulde02}, 
and enables many new aspects of these emergent excitations to be studied 
experimentally, such as pair creation and interaction, 
individual motion, currents of monopoles, correlations and cooperative phenomena. 
In the present study, inspired by the theoretical prediction 
of the monopoles, we have investigated two aspects of 
magnetic monopoles in spin ice using direct 
neutron scattering techniques and thermodynamic specific heat measurements.

Single crystals of Dy$_2$Ti$_2$O$_7$ were prepared by the floating-zone 
method \cite{Matsuhira_kagome}. 
Specific heat was measured by a quasi-adiabatic method. 
Neutron scattering experiments on a single crystal 
under a [111] field were performed on the triple-axis 
spectrometers BT-9 at the NIST Center for Neutron Research 
and the ISSP-GPTAS at the Japan Atomic Energy Agency. 
The sample was mounted in dilution refrigerators 
so as to measure the scattering plane perpendicular to the 
[111] direction.

\begin{figure}[tb]
\begin{center}
\includegraphics[width=7.0cm,clip]{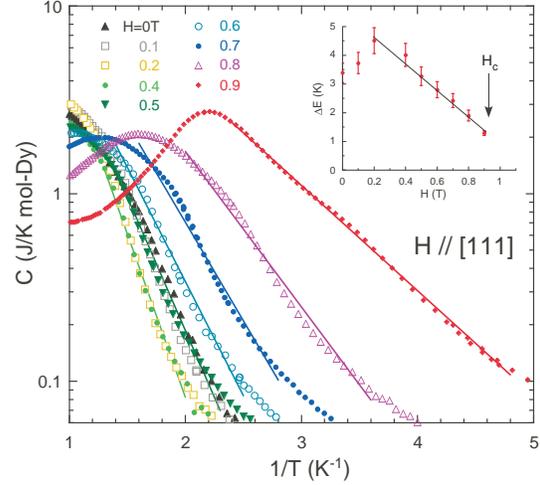}
\end{center}
\caption{
(Color) 
Specific heat under [111] fields 
is plotted as a function of $1/T$. 
In the intermediate temperature range these 
data are well represented by the Arrhenius law denoted 
by solid lines. 
The inset shows the field dependence of the activation energy.
}
\label{f2}
\end{figure}

A straightforward signature of monopole-pair creation is an 
Arrhenius law in the temperature ($T$) dependence of the 
specific heat ($C$), 
$C(T) \propto \exp(-\Delta E/k_{\mathrm {B}}T)$, where $\Delta E$ is a 
field ($H$) dependent creation energy. 
One can simply expect 
$\Delta E = E_0 - \mu H$ owing to the Zeeman effect. 
Figure~\ref{f2} shows the measured $C(T)$ of Dy$_2$Ti$_2$O$_7$ under a [111] 
applied field as a function of $1/T$. 
The Arrhenius law is clearly seen at low temperatures, 
indicating that monopole--antimonopole pairs are thermally 
activated from the ground state. 
We remark that all the measurements were performed under field cooling 
conditions, which are important to avoid complications due to spin 
freezing \cite{Bramwell01,Ramirez_DTO,melko_mc} among the ground 
state manifolds, whose degeneracy are slightly lifted. 
We think that the deviation from the Arrhenius law at the lowest 
temperatures is attributable to these perturbative effects. 
The observed activation energy $\Delta E$ depends linearly on 
$H$ (Fig.~\ref{f2} inset) between 0.2 and 0.9 T, i.e., in 
the kagom\'{e} ice state. 
The deviation from linearity below 0.2 T (spin ice regime) 
suggests that the nearest-neighbor effective bond energy 
$J_{\mathrm{eff}}$ slightly changes between the two states. 
The zero-field value $\Delta E(H=0)=3.5$ K reasonably agrees with an 
estimation $\Delta E(H=0)=4J_{\mathrm{eff}}=4.5$ K using $J_{\mathrm{eff}}$ 
in ref. 16. 
The observed Arrhenius law of $C(T)$, which is attributable to 
variation of density of the monopole pairs, implies that the 
number of monopoles can be tuned by changing $T$ and $H$.

A microscopic experimental method of observing monopoles is 
magnetic neutron scattering. 
One challenge to the experiments is to distinguish the relatively weak 
scattering from the small number of monopoles from the very strong magnetic 
scattering \cite{Bramwell01,Tabata_kagome} of the ground 
states. 
A theoretical idea \cite{Castelnovo08} which is helpful for identifying the 
monopole scattering is that the [111] field acts as chemical 
potential of the monopoles, enabling us to control their 
density as shown by the present specific heat measurements. 
As the field is increased, the kagom\'{e} ice state with low 
monopole density changes continuously to the 
maximum density state, the staggered monopole state (Fig.~\ref{f1}(e)), 
where all spin configurations become 
``3-in, 1-out'' or ``1-in, 3-out'' to minimize the Zeeman energy.

\begin{figure}[tb]
\begin{center}
\includegraphics[width=7.0cm,clip]{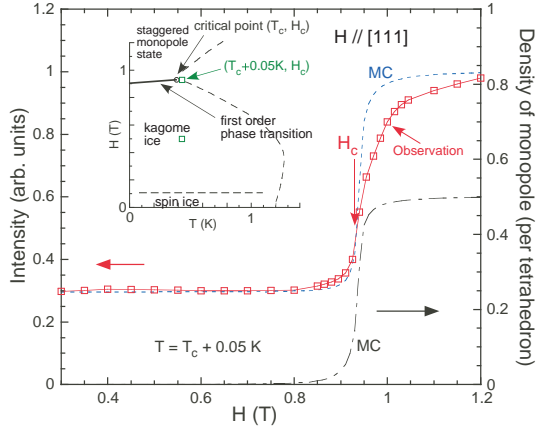}
\end{center}
\caption{
(Color) 
The magnetic Bragg intensity at $T=T_{\mathrm{c}}+0.05$ K is plotted 
as a function of the [111] field. 
The open squares and dashed curves represent the 
measurements at $(2 \bar{2} 0)$ and corresponding MC 
simulations, respectively. 
The dot-dashed curve is the density of positively charged 
monopoles obtained by the simulation. 
The inset shows the $HT$ phase diagram under the [111] field. 
The solid line represents the first-order phase transition 
with the critical point shown by an open circle \cite{sakakibara03}. 
The dashed lines are crossovers \cite{Higashinaka_kagome}. 
The intensity maps shown in Fig.~\ref{f4} were measured at the 
two points depicted by open squares.
}
\label{f3}
\end{figure}

For the present neutron scattering experiments, the best 
temperature and field region to observe monopoles in Dy$_2$Ti$_2$O$_7$ 
is close to the liquid-gas type critical 
point \cite{sakakibara03} $(T_{\mathrm{c}},H_{\mathrm{c}})$ (Fig.~\ref{f3} inset). 
In the monopole picture \cite{Castelnovo08}, where they are interacting via the 
magnetic Coulomb force, the first-order phase transition \cite{sakakibara03} 
is ascribed to phase separation between high- and low-density 
states. 
We naturally anticipate that neutron scattering close to the 
critical point is a superposition of the scattering pattern 
by the (low-density) kagom\'{e} ice state 
\cite{Tabata_kagome} and that by 
high-density monopoles, which is diffuse scattering around 
magnetic Bragg reflections, i.e., ferromagnetic fluctuations.

The neutron measurements were performed under a [111] 
field, and Monte Carlo (MC) simulations \cite{Tabata_kagome} 
were also carried out for the dipolar spin ice 
model \cite{Hertog_mc,melko_mc} to quantify our observations. 
Figure~\ref{f3} shows the field dependence of the magnetic intensity 
of the $(2 \bar{2} 0)$ Bragg reflection at $T=T_{\mathrm{c}}+0.05=0.43$ K. 
The intensity plateau for $H<0.8$ T corresponds to the kagom\'{e} ice 
state with low density monopoles. 
The deviation from the plateau as $H$ exceeds 0.8 T indicates 
that monopoles are being created gradually, 
while the saturation of the intensity for $H \gg H_{\mathrm{c}}$ denotes 
the staggered monopole state (Fig.~\ref{f1}(e)). 
In Fig.~\ref{f3} we also show the Bragg intensity and the density 
of the positively charged monopoles obtained by the simulation. 
The observation shows good agreement with the simulation 
for $H<H_{\mathrm{c}}$. 
On the other hand, above $H_{\mathrm{c}}$ there are substantially less monopoles 
than expected from the simulation, which will be discussed below.

\begin{figure*}[tb]
\begin{center}
\includegraphics[width=18.0cm,clip]{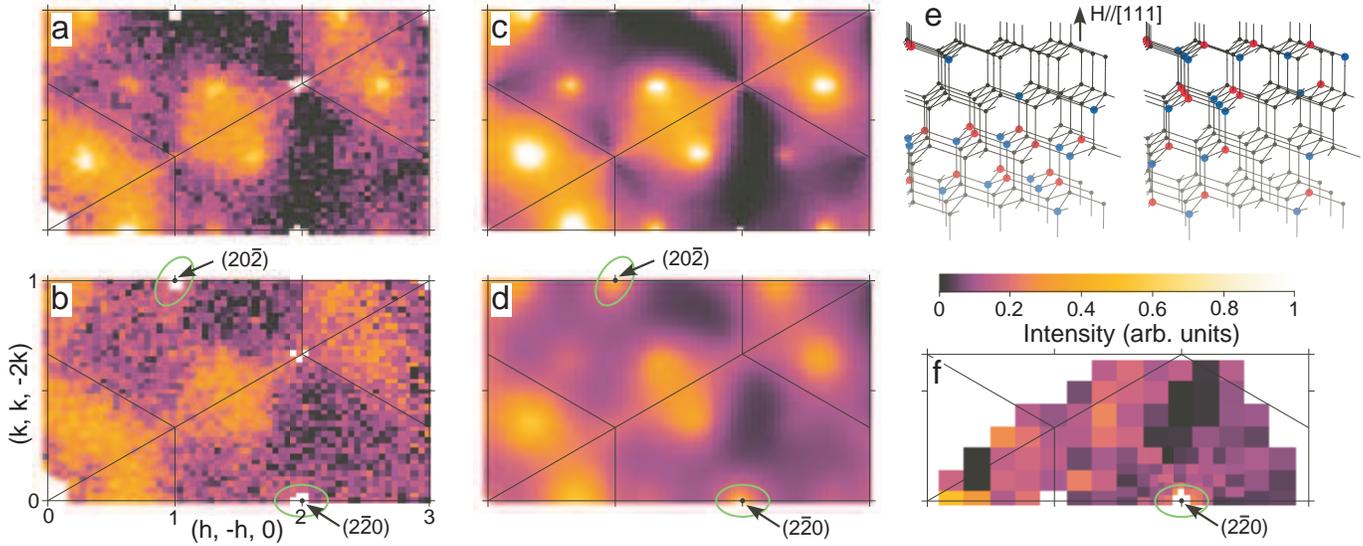}
\end{center}
\caption{
(Color) 
Intensity maps measured at $T=T_{\mathrm{c}}+0.05$ K 
in the scattering plane are shown for two 
field values (Fig.~\ref{f3} inset): 
the kagom\'{e} ice state at $0.5$ T (a, c) and the fluctuating 
high- and low-density monopoles at $H_{\mathrm{c}}$ (b, d, f). 
(a, b) and (f) were measured on the GPTAS and 
BT-9 spectrometers, respectively. 
(c, d) are simulated intensities. 
(e) Two snapshots of the monopoles of the simulation 
corresponding to (b, d, f) are shown on the diamond lattice, 
in which blue, red, and black points represents $+$, $-$ and 0 magnetic 
charges, respectively. 
The light green circles 
in (b, d, f) show the high intensity regions caused by the 
high-density monopoles, i.e., ferromagnetic fluctuations.
}
\label{f4}
\end{figure*}

We selected $T=T_{\mathrm{c}}+0.05$ K and 
$H=H_{\mathrm{c}}$ (Fig.~\ref{f3} inset) for observation of 
the fluctuating high- and low-density monopoles. 
At this $H$,$T$ point, we measured intensity maps in the scattering 
plane. 
An intensity map of the kagom\'{e} ice state at $T=T_{\mathrm{c}}+0.05$ K and 
$H=0.5$ T was also measured for comparison. 
Figure~\ref{f4} compares the measured and simulated intensity maps. 
The observed scattering pattern of the kagom\'{e} ice state 
(Fig.~\ref{f4}(a)) is in excellent agreement with the simulation 
(Fig.~\ref{f4}(c)), showing the peaked structure 
\cite{Tabata_kagome} at $(2/3, -2/3, 0)$ 
and the pinch point \cite{Fennel_kagome} at 
$(4/3, -2/3, -2/3)$. 
These structures reflect the kagom\'{e} ice state.

The observed (Fig.~\ref{f4}(b)) and simulated (Fig.~\ref{f4}(d)) intensity maps 
close to the critical point show a weakened kagom\'{e}-ice scattering 
pattern (by the low-density state) and diffuse scattering 
around $(2 \bar{2} 0)$ (by the high-density state). 
The observation agrees fairly well with the simulation. 
However the diffuse scattering is less pronounced for the observation. 
We found that this discrepancy originated from an instrumental 
condition of the GPTAS spectrometer, which has a large vertical resolution of 
$\Delta q=0.25$ {\AA}$^{-1}$ (full width at half maximum, FWHM). 
We carried out the same measurement on the BT-9 spectrometer. 
It has a smaller vertical resolution of $\Delta q=0.1$ {\AA}$^{-1}$ (FWHM), 
which does not affect the diffuse scattering. 
The resulting data are shown in Fig.~\ref{f4}(f), which are in better agreement 
with the simulation (Fig.~\ref{f4}(d)) around $(2 \bar{2} 0)$. 
An interesting point suggested by this resolution 
effect is that correlations of the high-density monopoles are 
three dimensional in space, 
although the monopoles can only move in the two dimensional 
layers (Fig.~\ref{f1}(d)). 
The three dimensional correlations are consistent with 
the isotropic Coulomb interaction between monopoles. 
We note that the kagom\'{e}-ice scattering pattern is two 
dimensional in nature \cite{Tabata_kagome}, and thus is not affected by 
the vertical resolution.

To illustrate the high- and low-density monopoles yielding 
the scattering patterns in Figs.~\ref{f4}(b), 4(d), and 4(f), two typical 
snapshots of the monopoles of the MC simulation are shown 
in Fig.~\ref{f4}(e), where we depict magnetic charges at centers of 
the tetrahedra. 
Lines connecting the centers of tetrahedra form a diamond 
lattice, and magnetic charges reside on its lattice points. 
Regions of the low-density monopoles (where black points 
dominate) produce the kagom\'{e}-ice scattering pattern, 
while those of the high-density monopoles (where red 
and blue points dominate) produce the diffuse 
scattering around the Bragg reflections. 
These critical fluctuations between high- and low-density 
phases reinforce the proposed explanation \cite{Castelnovo08} of the 
puzzling liquid-gas type critical point \cite{sakakibara03} using the 
similarity argument to phase transitions of ionic particle 
systems on lattices \cite{Kobelev_02}. 
Consequently they strongly suggest existence of magnetic monopoles 
interacting via the magnetic Coulomb force. 
Further investigations of 
critical phenomena, \cite{Higashinaka_kagome} screening of 
the Coulomb interaction, and effects of the anisotropic 
motion of the monopoles within the kagom\'{e} lattices are of interest.

A question, which is not pursued in the present study, 
is how monopoles unbound by the fractionalization move 
in the kagom\'{e} lattice. 
Comparing the observed Arrhenius law with that of a 
study \cite{Jaubert_09} 
of the diffusive motion of monopoles in spin ice state, 
it seems that the interesting temperature ranges where 
unconfined monopoles move diffusively are roughly 
$T>1.5$ K ($H=0.5$ T) and $T>0.7$ K ($H=0.9$ T).
There may be another interesting issue in the discrepancy between 
observed Bragg intensity and the classical MC 
simulation shown in Fig.~\ref{f3} ($H > H_{\mathrm{c}}$). 
We speculate that it may indicate the existence of quantum mechanical 
effects neglected in the computation. 
Puzzling experimental facts at low $T$ were also noticed by 
the slow saturation of magnetization \cite{sakakibara03} 
and the non-zero specific heat \cite{Higashinaka_kagome} 
above $H_{\mathrm{c}}$ down to very low temperatures, 
$T < 0.1$ K. 
For example, if the double spin flips shown in Fig.~\ref{f1}(d) 
can occur by tunneling \cite{Ehlers_03}, monopoles (or holes 
in the staggered monopole state) might move more easily 
than classical diffusion \cite{Jaubert_09}. 

Typical elementary excitations in condensed matter, such as 
acoustic phonons and magnons, are Nambu-Goldstone 
modes where a continuous symmetry is spontaneously broken 
when the ordered state is formed. 
This contrasts with the monopoles in spin ice, which are 
point defects that can be fractionalized in the frustrated 
ground states. 
Such excitations are unprecedented in condensed matter, 
and will now enable conceptually new emergent phenomena 
to be explored experimentally.


\begin{thebibliography}{99} 

\bibitem{Jackson}
J. D. Jackson: 
{\it Classical Electrodynamics} 
(Wiley, New York, 1975) Chap. 6.12-6.13. 

\bibitem{Dirac31}
P. A. M. Dirac: 
Proc. R. Soc. A {\bf 133} (1931) 60.

\bibitem{Milton06}
K. A. Milton: 
Rep. Prog. Phys. {\bf 69} (2006) 1637. 

\bibitem{Fang03}
%Z. Fang {\it et al.}: 
Z. Fang, N. Nagaosa, K. S. Takahashi, A. Asamitsu, 
R. Mathieu, T. Ogasawara, H. Yamada, M. Kawasaki, 
Y. Tokura, and K. Terakura: 
Science {\bf 302} (2003) 92.

\bibitem{Castelnovo08}
C. Castelnovo, R. Moessner, and S. L. Sondhi: 
Nature {\bf 451} (2008) 42. 

\bibitem{Qi09}
X-L. Qi, R. Li, J. Zang, and S-C. Zhang: 
Science {\bf 323} (2009) 1184.

\bibitem{Harris_HTO}
M. J. Harris, S. T. Bramwell, D. F. McMorrow, T. Zeiske, and K. W. Godfrey: 
Phys. Rev. Lett. {\bf 79} (1997) 2554. 

\bibitem{Bramwell01}
S. T. Bramwell and M. J. P. Gingras: 
Science {\bf 294} (2001) 1495.

\bibitem{Pauling1935}
L. Pauling: 
J. Am. Chem. Soc. {\bf 57} (1935) 2680.

\bibitem{Ramirez_DTO}
A. P. Ramirez, A. Hayashi, R. J. Cava, R. Siddharthan, and B. S. Shastry: 
Nature {\bf 399} (1999) 333.

\bibitem{Matsuhira_kagome}
K. Matsuhira, Z. Hiroi, T. Tayama, S. Takagi, and S. Sakakibara: 
J. Phys. Condens. Matter {\bf 14} (2002) L559.

\bibitem{Moessner03}
R. Moessner and S. L. Sondhi: 
Phys. Rev. B {\bf 68} (2003) 064411.

\bibitem{sakakibara03}
T. Sakakibara, T. Tayama, Z. Hiroi, K. Matsuhira, and S. Takagi: 
Phys. Rev. Lett. {\bf 90} (2003) 207205.

\bibitem{Tabata_kagome}
Y. Tabata, H. Kadowaki, K. Matsuhira, Z. Hiroi, N. Aso, E. Ressouche, and B. F{\aa}k: 
Phys. Rev. Lett. {\bf 97} (2006) 257205.

\bibitem{Higashinaka_kagome}
R. Higashinaka, H. Fukazawa, K. Deguchi, and Y. Maeno: 
J. Phys. Soc. Jpn. {\bf 73} (2004) 2845.

\bibitem{Hertog_mc}
B. C. den Hertog and M. J. P. Gingras: 
Phys. Rev. Lett. {\bf 84} (2000) 3430.

\bibitem{Fulde02}
P. Fulde, K. Penc, and N. Shannon: 
Ann. Phys. {\bf 11} (2002) 892.

\bibitem{melko_mc}
R. G. Melko and M. J. P. Gingras: 
J. Phys. Condens. Matter {\bf 16} (2004) R1277.

\bibitem{Fennel_kagome}
T. Fennell, S. T. Bramwell, D. F. McMorrow, P. Manuel, and A. R. Wildes: 
Nature Physics {\bf 3} (2007) 566.

\bibitem{Kobelev_02}
V. Kobelev, A. B. Kolomeisky, and M. E. Fisher: 
J. Chem. Phys. {\bf 116} (2002) 7589.

\bibitem{Jaubert_09}
L. D. C. Jaubert and P. C. W. Holdsworth: 
Nature Physics {\bf 5} (2009) 258.

\bibitem{Ehlers_03}
%G. Ehlers {\it et al.}: 
G. Ehlers, A. L. Cornelius, M. Orend{\'a}c, M. Kajnakov{\'a}, T. Fennell, 
S. T. Bramwell, and J. S. Gardner: 
J. Phys. Condens. Matter {\bf 15} (2003) L9.

\end{thebibliography}
\end{document}